\documentclass[11pt]{article}
\usepackage{moriond}
\usepackage{graphicx}
\usepackage{array}
\usepackage{enumitem}
\usepackage{changepage}

\def\be{\begin{equation}}
\def\ee{\end{equation}}
\def\bea{\begin{eqnarray}}
\def\eea{\end{eqnarray}}

\newcommand{\Photo}{\includegraphics[height=35mm, keepaspectratio]{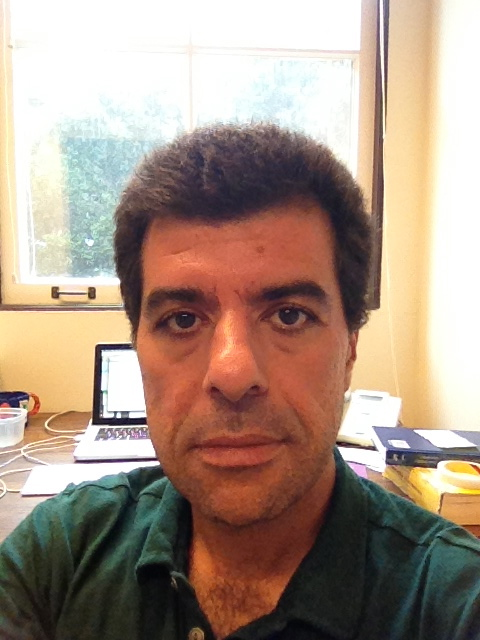}}

\begin{document}
\vspace*{4cm}
\title{Dark Matter Direct Detection With Low Temperature Detectors}

\author{ N.MIRABOLFATHI }

\address{Department of Physics, University of California, Berkeley, 343 Leconte Hall, \\UC Campus, 
Berkeley, CA 94720 USA}

\maketitle\abstracts{  Low Temperature Detectors (LTDs) offer the best signal-to-noise performance and thus, lowest energy threshold among other particle detection technologies available today. The excellent background rejection provided by these detectors is essential to search for very rare and very low energy recoils expected from dark matter interactions. Furthermore, due to their very good event-by-event background rejection capabilities, LTDs do not heavily rely on self-shielding in order to mitigate radioactive backgrounds (in contrast with the monolithic liquid Noble Gas detectors) thus they can be more easily prototyped and tested for basic detector performance. A review of the status and prospects of the major Dark Matter search experiments using LTD technologies is presented here.}

\section{Dark Matter and WIMPs}

A broad range of cosmological observations, including those of the distribution of large-scale structure,
of the apparent brightness of distant supernovae, and of the anisotropies in the cosmic microwave
background, tells us that about 85\% of the matter in the universe is not made of ordinary
particles, but exists in some dark form \cite{Planck}.

Weakly Interacting Massive Particles (WIMPs) are a generic class of candidates for dark matter~\cite{lee1977}$^{,}$\cite {jungman}. As Big Bang relic particles, WIMPs are particularly interesting because their existence would represent a convergence of independent arguments from cosmology and particle physics. They constitute a class of  particles  produced in the hot early universe in thermal equilibrium that decoupled when they were non-relativistic. In order for them to have a density equal to that of dark matter,  they should interact with a cross section similar to that of the Weak Interaction (hence their name). Therefore, cosmology hints at the possibility that physics at the scale of the W and Z bosons may be responsible for dark matter. In particle physics, new physics also appears to be needed at the W and Z scale in order to solve the famous ``hierarchy'' problem: We need to understand why m$_{H}\sim$125 GeV/c$^{2}$  in spite of the radiative corrections that tend to drive its mass to high values. Supersymmetry is the primary example of how such radiative corrections can be controlled, and (if R-parity is conserved) the lightest supersymmetric partner ({\small LSP}) is stable and may interact at roughly the weak interaction rate. Accordingly, such models predict a relic density of the {\small LSP} very near the universe's inferred dark matter density. 

\begin{figure*}[htbp]
\vspace{-10pt}
\begin{center}
\centering
\includegraphics[  width=0.75\linewidth,  keepaspectratio]{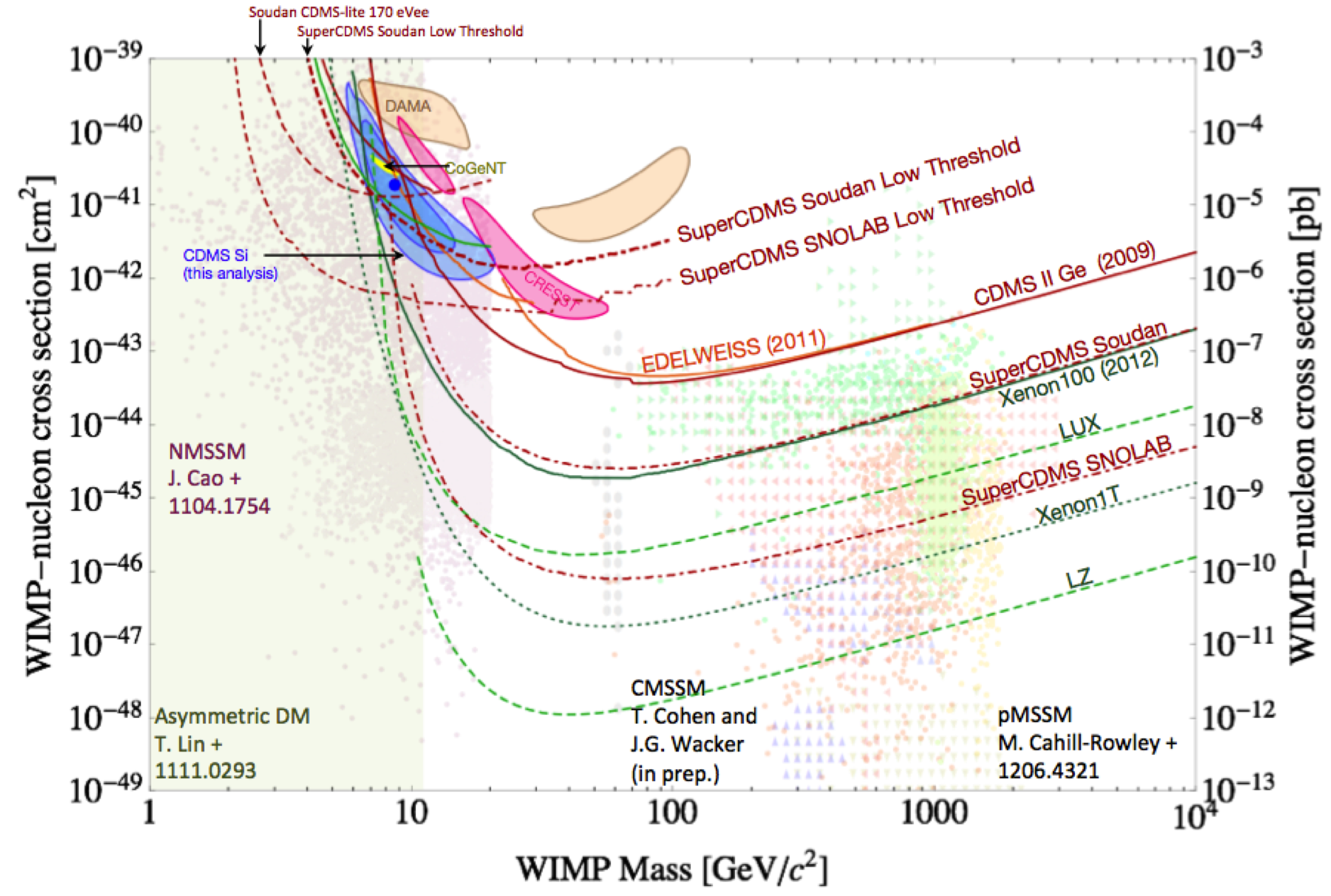}
\caption{\footnotesize Recent upper limits (90\% C.L.) on the WIMP-nucleon spin-independent cross section versus~\protect\cite{CDMSIISi}
WIMP mass are shown from top to bottom for CDMS~II~Soudan\protect\cite{SoudanLowthresh2011}$^{,}$~
\protect\cite{c58science} (Red solid), EDELWEISS~II~\protect\cite{Edelweiss2011} (Orange solid) and XENON100~
\protect\cite{Xenon100final} (green solid). The magenta filled region indicates the region where CRESST II reports a signal~
\protect\cite{CRESSTII730kgd}. 68$\%$ and 90 $\%$ C.L. contours for a possible signal from CDMS~II recent silicon 
analysis~\protect\cite{CDMSIISi} are shown in light blue. The blue dot shows the maximum likelihood point at (8.6 GeV/c
$^{2}$, 1.9$\times$10$^{-41}$ cm$^{2}$) The colored regions show the  current SUSY regions (with recent LHC and 
Higgs constraints). Also shown are projected sensitivities for the SuperCDMS Soudan experiment (dot-dashed red) 
and the proposed SuperCDMS SNOLAB experiment with a 200\,kg payload for four years running (dot-dashed red); 
these assume no background subtraction. Low-threshold SuperCDMS Soudan (dashed red) and SuperCDMS 
SNOLAB (dashed red) projections are also shown.
}
\label{fig:SILimPlot}
\end{center}
\vspace{-10pt}
\end{figure*}

Fig.~\ref{fig:SILimPlot} shows the present experimental situation together with a range of supersymmetric models, for ``spin-independent'' scattering, where the quantum number that adds in the coherent scattering matrix element is assumed to be the atomic number of the nucleus.

There are two regions of current interest: the mass region above 50\,GeV/c$^2$  favored by most WIMP models, and the low-mass region around 7\,GeV/c$^2$. In the higher mass region, the best upper limits are given by three experiments, XENON100~\cite{Xenon100final}, CDMS~II~\cite{c58science}, and EDELWEISS~\cite{Edelweiss2011} 
These limits exclude parts of the supersymmetry parameter space allowed by the recent LHC results~\cite{cMSSM}. 
Although the cMSSM/mSUGRA 5-parameter space is severely restricted by the LHC, more general supersymmetric models are fully compatible with the current data. Generally speaking, the absence of missing energy events at LHC tends to push the mass scales higher. It is important to note that direct detection has no sharp high-mass cutoff as does the LHC; instead the sensitivity merely degrades as the inverse of the WIMP mass.

The low-mass WIMP region shown in Fig.~\ref{fig:SILimPlot} is interesting in part due to the experimental claims by DAMA~\cite{HooperPRD2010} (based on modulation assuming Na scattering), CoGeNT~\cite{CoGeNT} (based on differential rate and modulation) and CRESST~\cite{CRESSTII730kgd} and more recently the unexpected excess in the CDMS~II silicon analysis~\cite{CDMSIISi}.
This mass region is also interesting from the purely theoretical point-of-view because it is natural in asymmetric dark matter models, with a dark-matter anti-dark-matter asymmetry of the same order of magnitude as in the baryonic sector~\cite{AsymmetricDM}. If the annihilation rate is large enough for the pairs of particles and antiparticles to disappear, then the amount of dark matter would be readily explained by the ratio of masses, and a relatively high elastic cross section could be natural. 

\section{Direct detection of WIMPs challenges: LTD advantages }

If WIMPs are indeed the main components of dark matter and form a halo around our galaxy, then they can interact with a terrestrial detector leaving a detectable elastic scattering off the target nuclei~\cite{Gaitskell review}. The rate of such interaction depends on the local halo density and velocity distribution in the Milky Way, the WIMP mass, and the cross section on the target nuclei. The last of these parameters has the largest uncertainty.  Generally, the recoil energy spectrum is given by
\vspace{-5pt}
\begin{equation}
\frac{dN}{dR}=\frac{\sigma_0\rho_\chi}{2\mu^2m_\chi}F^2(q)\int_{\upsilon_{min}}^{\upsilon_{escape}}\frac{f(\upsilon)}{\upsilon}d\upsilon,
\label{eq:rates}
\end{equation}

where $\rho_\chi$ is the local WIMP density,$ \mu$ is the WIMP-nucleus reduced mass
$m_\chi m_N /(m_\chi + m_N )$ (assuming a target nucleus mass $m_N$ ), and the integral takes
account of the velocity distribution $f (\upsilon)$ of WIMPs in the halo. The term $v_{min}$ is
the minimum WIMP velocity able to generate a recoil energy of $E_r$ , and $v_{esc}$ is the
maximum WIMP velocity set by the escape velocity in the halo model. $F^{2}(q)$ is
the nuclear form factor and $\sigma_{0}$ the WIMP nucleus interaction cross section.

\begin{figure*}
\vspace{-10pt}
\begin{center}
\centering
\includegraphics[width=0.45\linewidth,  keepaspectratio]{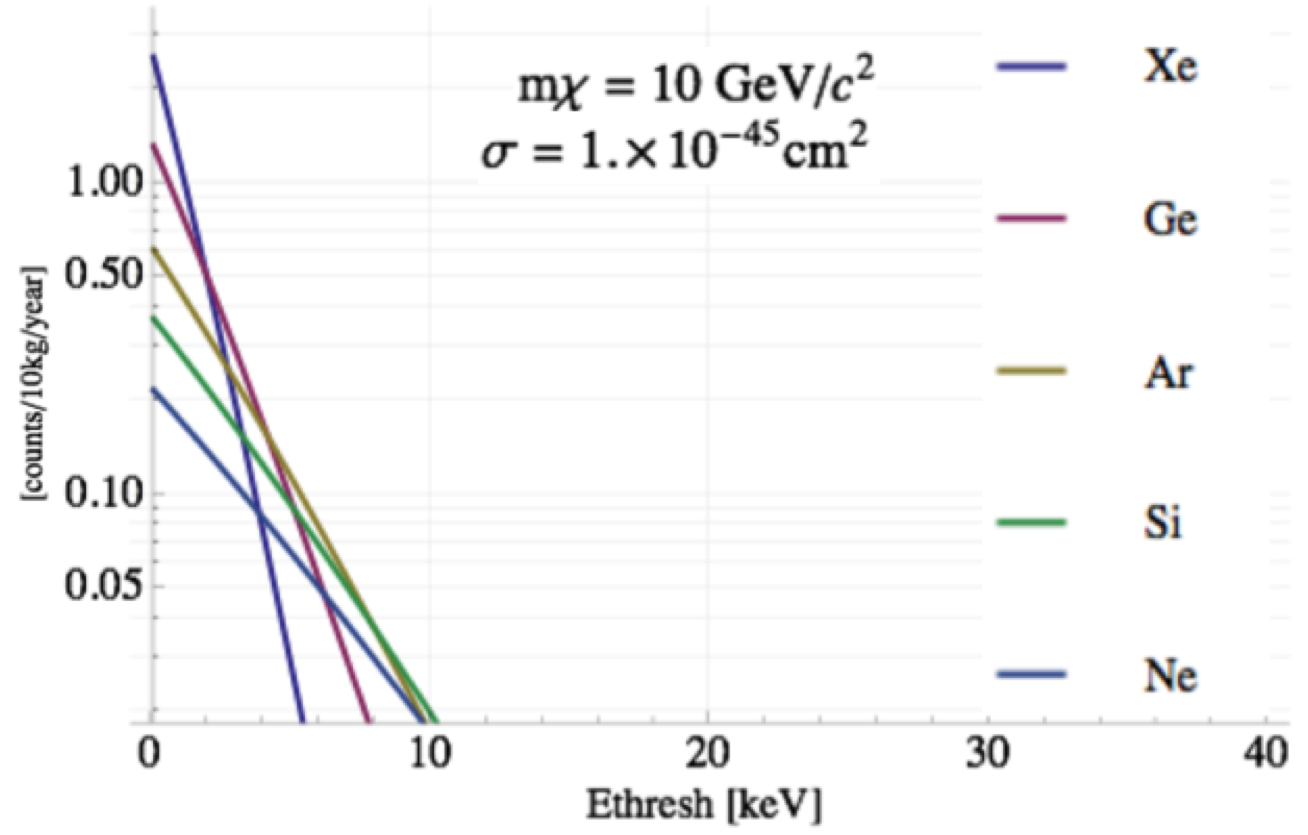}
\includegraphics[width=0.45\linewidth,  keepaspectratio]{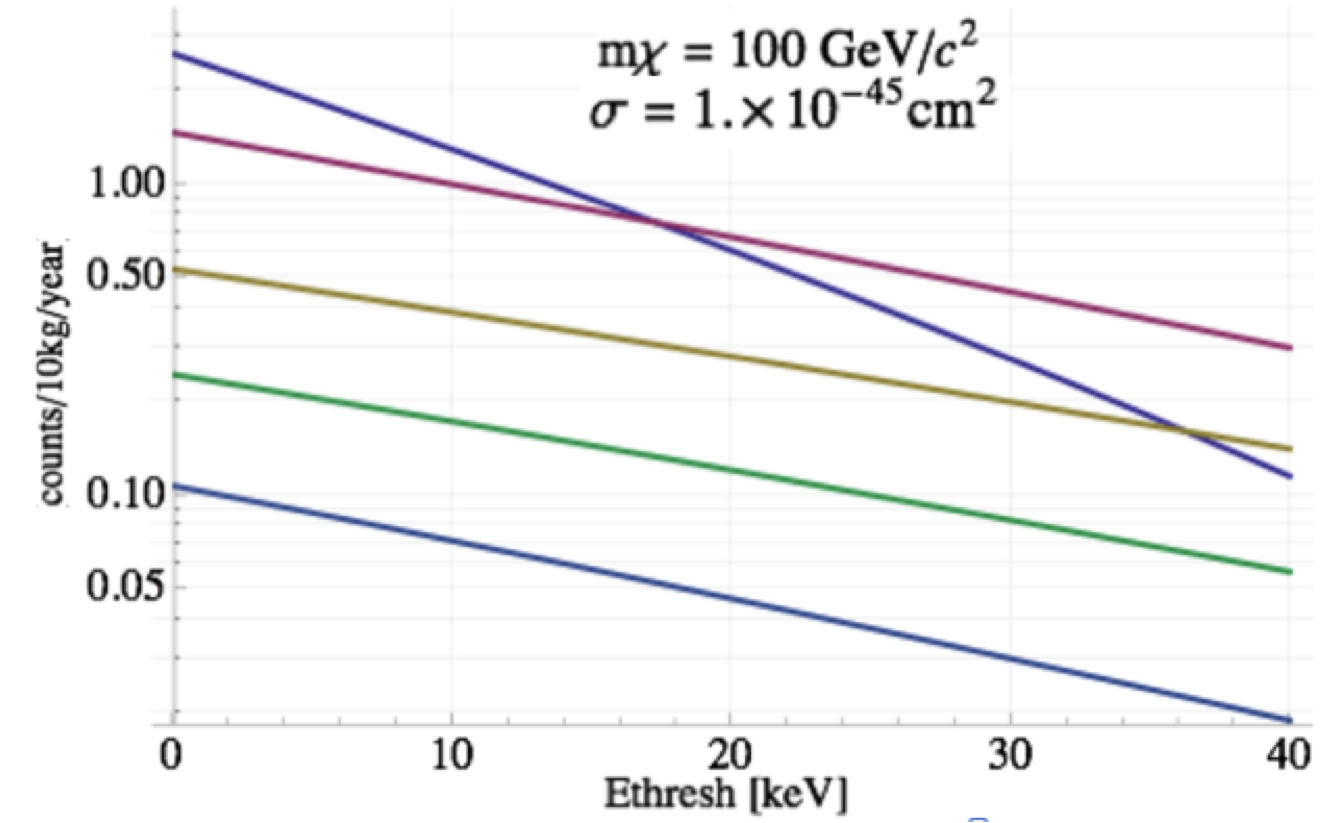}
\caption{\footnotesize The total event rate above detections threshold in various substrates for a WIMP mass$=100 GeV/c^{2}$ (left) and WIMP mass$=10 GeV/c^{2}$ (right) for a cross-section of interaction= $1\times10^{-45}cm^{2}$.}
\label{fig:Interaction rates}
\end{center}
\vspace{-10pt}
\end{figure*}

Fig.~\ref{fig:Interaction rates}, shows the expected interaction rates for various detector material and WIMP masses of: 100 GeV/c$^{2}$ (left) and 10 
GeV/c$^{2}$ (right). These plots also reveal the challenging aspects of WIMP direct detection notably:

\begin{itemize}
\item \emph{ Low energy recoils}  require low detection thresholds and high resolution detectors.
\item \emph{ The expected recoil spectrum} is a featureless, simple falling exponential.
\item \emph{Very rare interactions} requires background mitigation.
\end{itemize}

Detectors operating below a temperature of 1 K, also as known as Low Temperature Detectors (LTDs)~\cite{Cryo book}$^{,}$\cite{LTD14}, use $<$meV quanta phonons to provide better energy resolution than is typically available from conventional technologies. Additionally the fact that all the quantum excitations induced by particle interaction will decay to the lowest excited states (phonons), the measurement of phonons provides a simple parameter to estimate the energy of interaction i.e. absolute calorimetry. The low energy threshold offered by LTDs is particularly important for light WIMP detection low threshold requirement. 

The most basic kind of LTD employs a dielectric absorber coupled to a thermal bath via weak link. A thermistor monitors the energy of the absorber. The energy $E$ deposited by a particle interaction causes calorimetric temperature change by increasing the population of thermal phonons. The fundamental sensitivity is
\vspace{-5pt} 
\begin{equation}
\sigma_{E}^{2}={\xi}^{2}kT(TC(T)+{\beta}E)
\label{Phonon sensitivity}
\end{equation}

where $C$ is the heat capacity of the detector and $T$ is the temperature of operation, k is Boltzmann's constant, $\xi$ and $\beta$ are parameters that depend on the nature of thermal link and other noise sources in the readout and are both of the order of unity. The first term represents phonon number fluctuations in the absorber due to thermal exchange with the bath, while the second term represents fluctuations in the number of phonons excited by particle interaction. Eq.~\ref{Phonon sensitivity} suggests the use of crystalline dielectric absorbers at very low temperatures particularly because the heat capacity of the dielectrics drops as T$^{3}$ for the temperatures well below the Debye temperature. ($\Theta_D$, typically hundreds of K). A $\sigma_{E}=5.2 eV$ is in principle achievable for 1 kg of germanium operated at $T$=10 mK. In practice, a number of factors ($e.g.$ readout noise) degrade the energy resolution by about an order of magnitude, but the fundamental energy resolution for such a large mass remains attractive. There are some commonly used sensors for LTDs. For example, Neutron-Transmutation-Doped (NTD) sensors have an exponentially decreasing resistance with temperature, while Transition Edge Sensors (TES) have a strong resistance change at their critical temperature, T$_{c}$.


In the absence of backgrounds, the sensitivity of an experiment to WIMP interaction increases linearly with the mass of the detector, which is an incentive for the desire to have more massive detectors. However, the thermal phonon measurement resolution degrades with the mass (or heat capacity) of the absorber as $\sqrt{M}$. This motivates the use of athermal phonons, $i.e.$ phonons which have not reached thermal equilibrium. Through successive anharmonic decays, the high energy phonons from the recoil rapidly decay into lower energy phonons.  Since these lower energy phonons have a mean free path comparable to the physical size of the detector, they travel ballistically from the interaction point to the surface of the detector and the phonon sensors therein. Assuming that the sensor area coverage on the surfaces of the detector increases commensurate to the detector mass, one can maintain the same energy resolution independent of the mass of the absorber. Furthermore, since the athermal phonons carry the information about the transient in the absorber, one can also use phonon signals to reconstruct  location of the events.

 Measurement of a single parameter per interaction ($e.g.$ phonon amplitude) does not provide any background discrimination capability. For an event-by-event discrimination one needs at least two simultaneous measurements. Ionization or scintillation, which are among the other standard nuclear and particle detection methods at higher temperatures, can also be measured at very low temperatures. They can be combined with a phonon measurement technique to discriminate WIMP producing Nuclear Recoils (NR) from most of the backgrounds producing Electron Recoils (ER). In particular, it has been shown that the ionization or scintillation yield of an interaction can vary significantly between a NR and an ER~\cite{c58science}$^{,}$~\cite{CRESSTII730kgd}. A brief summary of the leading WIMP search experiments using various LTD technologies is presented in Table~\ref{tab:Experiments summary} and more details are discussed in the following section. 

\begin{center}
\begin{table}[tp]
\caption{Summary of the major dark matter search experiments using LTDs. The third column shows results from neutron source calibration (to mimic WIMPs) from corresponding experiments. The Nuclear Recoils (NR) events centered at the lower scintillation or ionization yields and Electron recoil (ER) events centered around the yield of 1 are well separated resulting in excellent ER background discrimination.}\label{tab:Experiments summary}

 \begin{adjustwidth}{-1cm}{}
  \begin{tabular}[c]{|m{2cm} | m{3cm} | m{5cm} | m{6cm} | }
    \hline
    Experiment&Detector & Example of Calibration & Status and Perspective \\ \hline 
  CRESST&\includegraphics[width=30mm, keepaspectratio]{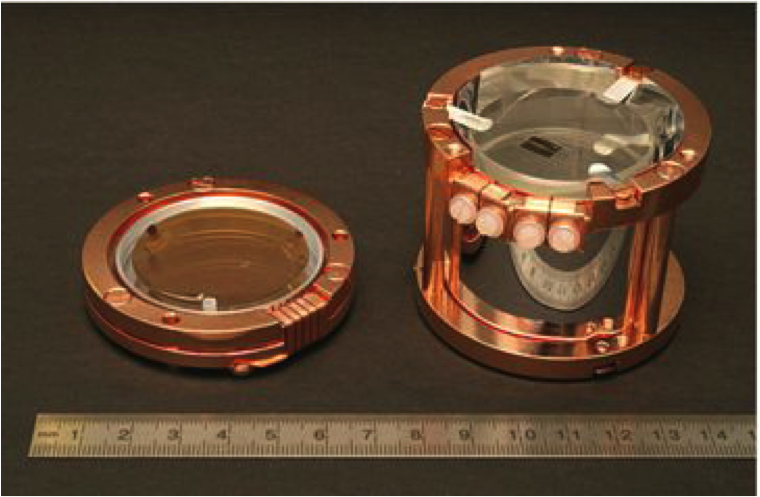} Scintillation and Phonon in 300~g CaWO$_{4}$ modules &\includegraphics[width=50mm, keepaspectratio]{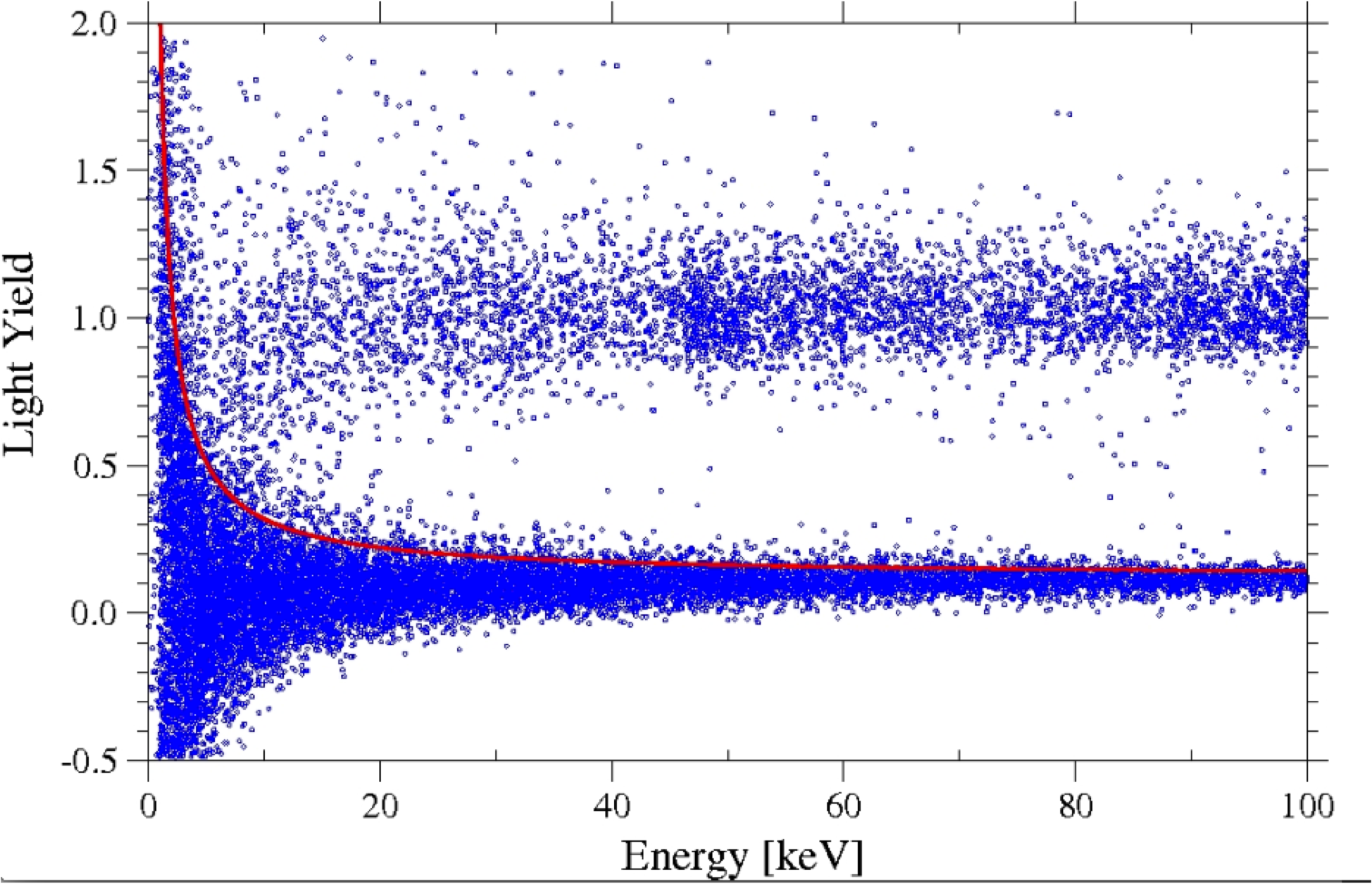}&\begin{itemize}\item Observed 67 candidates in 730 kg.day data.\item More than half are from backgrounds. \item CRESST II running: 5 kg detectors.\item Will confirm or constrain results from the previous run.\end{itemize}
       \  
    \\ \hline
EDELWEISS&\includegraphics[width=30mm, keepaspectratio]{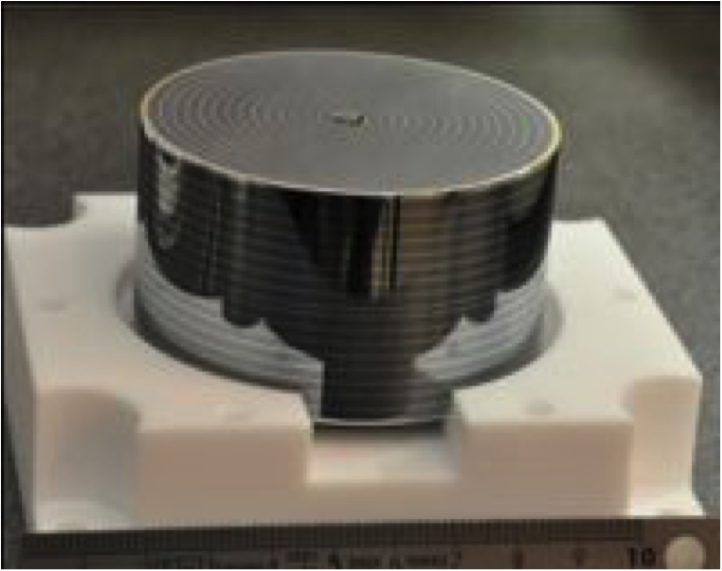} Ionization and thermal Phonon in 400 or 800 g Ge IDs or FIDs&\includegraphics[width=50mm, keepaspectratio]{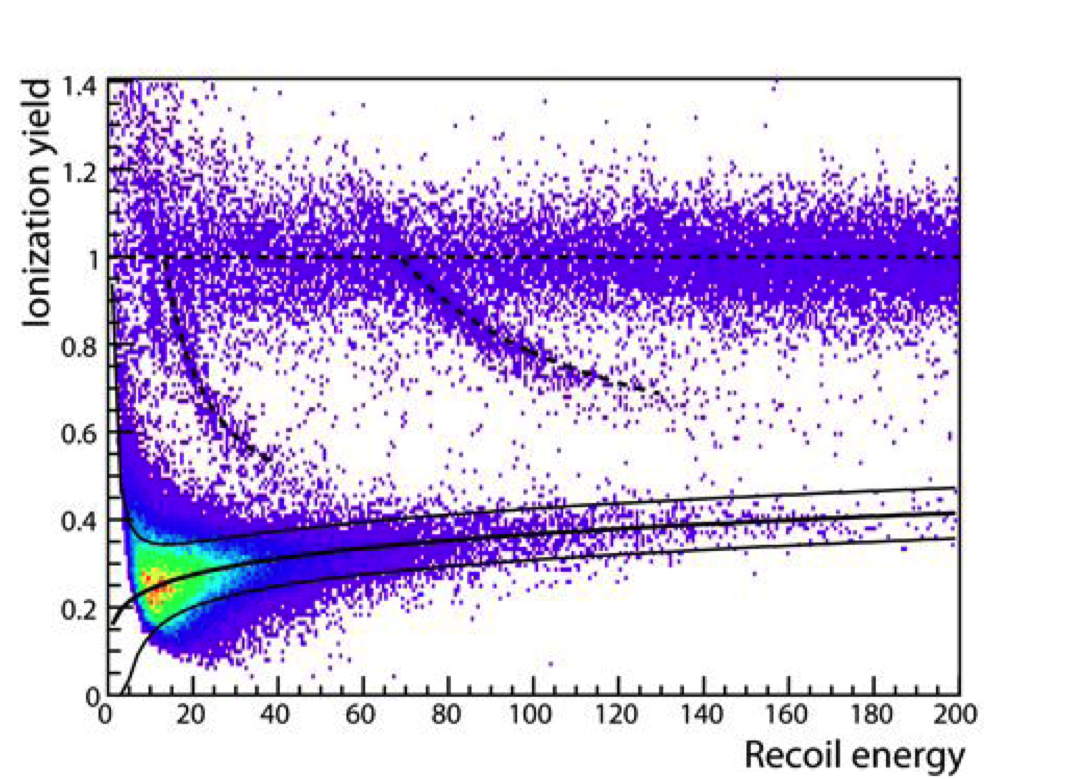}&\begin{itemize}\item EDELWEISSII exposure:  384 kg.day [20-100 keV] and 113 kg.day [5-20] keV\item EDELWEISS II sensitivity: $\sigma_{SI}<4.4\times10^{-44} cm^{2}$ ($\sigma_{SI}<1\times10^{-41} cm^{2}$ ) For 85 Gev/c$^{2}$ (10 Gev/c$^{2}$ ) WIMPs.\item EDELWEISS III will soon run with 40 kg FIDs\end{itemize}
       \  
    \\ \hline
CDMS&\includegraphics[width=30mm, keepaspectratio]{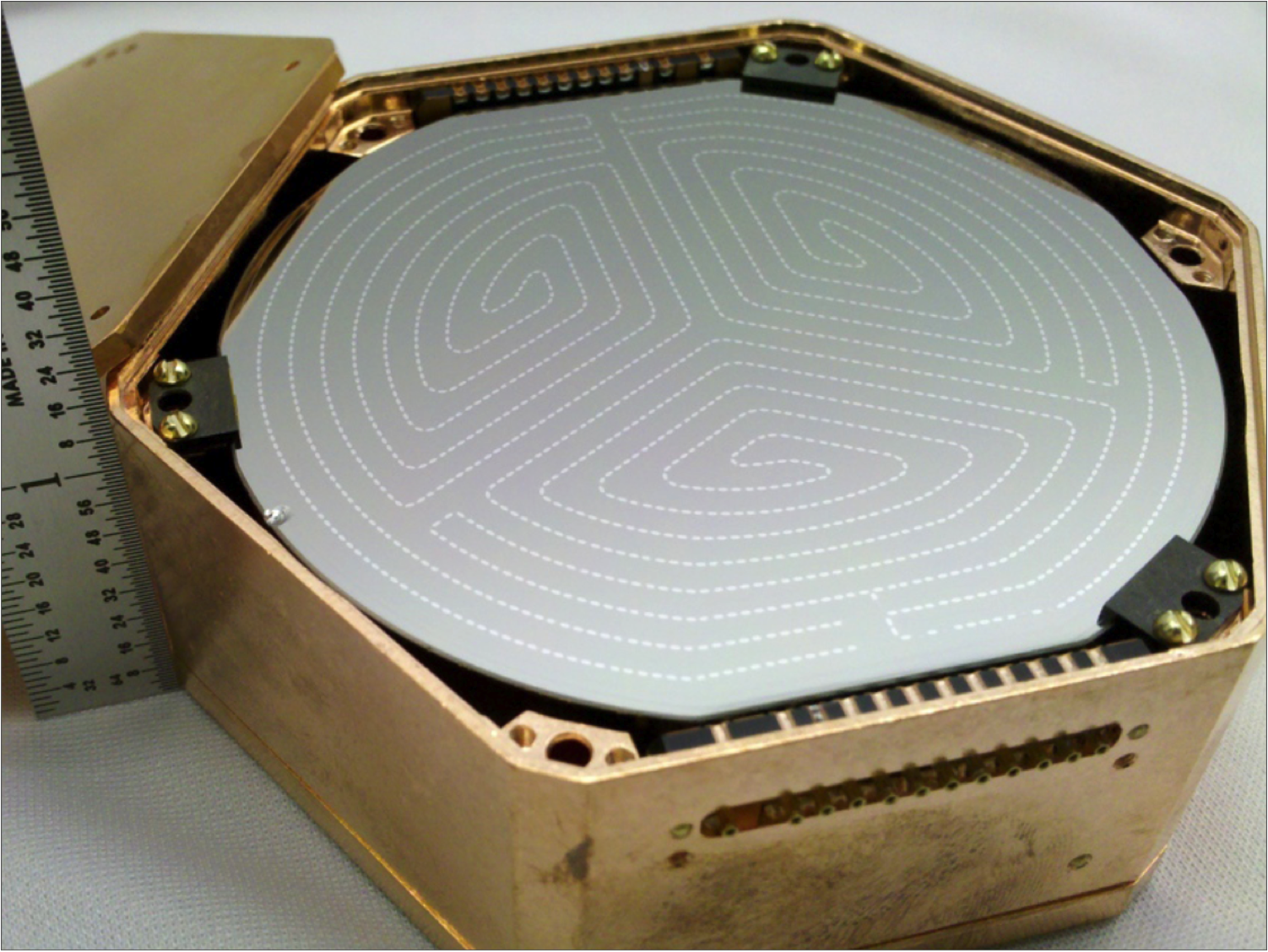} Ionization and athermal Phonon in 0.6 or 1.5 kg Ge iZIPs&\includegraphics[width=50mm, keepaspectratio]{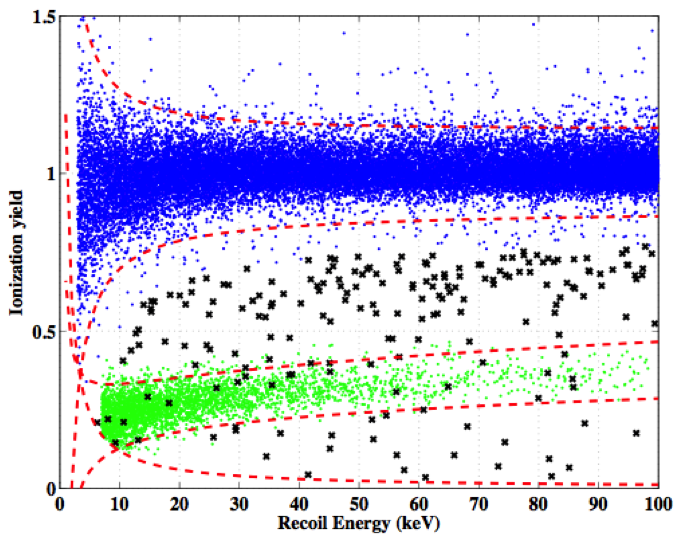}& \begin{itemize}\item CDMSII sensitivity with 4.5 kg Ge:  $\sigma_{SI}<4.4\times10^{-44} cm^{2}$ for 85 Gev/c$^{2}$ WIMPs\item CDMSII Observed 3 candidates in 142 kg.day Si data. with sensitivity: $\sigma_{SI}=2.4\times10^{-41} cm^{2}$ for 10 Gev/c$^{2}$ WIMPs\item SuperCDMS Soudan currently running with $\sim$ 9 kg iZIPs \item SuperCDMS SNOLAB R\&D for 200 kg of large 1.5 kg iZIPs
\end{itemize}
       \  
    \\ \hline
\end{tabular}
\end{adjustwidth}
\end{table}
 \end{center}

\section{Status and prospects of  direct detection experiment using LTDs}\label{subsec:prod}

\textbf{Scintillation}-The CRESST II experiment (Cryogenic Rare Event Search using Superconducting Thermometers) uses crystals of CaWO$_{4}$  as dark matter detectors~\cite{CRESSTII730kgd}. Each detector module is 300 g and can simultaneously detect scintillation and phonons for each event. This enables the experiment to discriminate WIMPs from radioactive backgrounds using the scintillation yield as shown on table.~\ref{tab:Experiments summary}. The detectors are operating at T$_{base}\sim$10 mK. The total recoil energy of the interaction is measured thermally with a Superconducting Phase Transition (SPT) thermometer attached to the detector. The photons are measured by a separate light detector placed immediately above the CaWO$_{4}$ crystal. Photons absorbed by the light detector lead to a temperature change that is measured by a second SPT. Using this technique the CRESST II experiment has established a background rejection efficiency of $>$99.7\% above 15 keV.
As a multi-nucleus target, CaWO4 crystals produce a scintillation yield that varies according to the type of recoiling nucleus. This allows CRESST II to probe different WIMP mass scenarios in the same target. However, this requires  well defined band definition down to the recoil energy of the interest below 20 keV.
After accumulating 730 kg.day of data the CRESST~II experiment observed 67 events within the WIMP signal region of  which roughly half could be interpreted as WIMPs of the mass 12 to 30 $GeV/c^{2}$. No strong claim is made of a WIMP discovery due to the unaccounted systematics associated with the known backgrounds. The experiment is currently running with $\sim$ 5 kg of crystals and with improved background environment to either confirm the low mass WIMPs detection in the previous run or produce competitive limits over a wide range of WIMPs masses.

\textbf{Ioniozation} in semiconductors in the form of electron-hole excitations can be measured at low temperatures. 
As shown in table.~\ref{tab:Experiments summary}, the ionization yield varies strongly between an electron recoil  and a nuclear recoil event. This provides a powerful handle to discriminate WIMP interactions from the radioactive background. Both CDMS (Cryogenic Dark Matter Search) and EDELWEISS (Experience pour DEtecter Les Wimps En Site Souterrain) use simultaneous measurement of Ionization and phonons for background discrimination.  

The phonon signals have two components: Phonons produced immediately after the recoils and those produced during the drift of scattering carriers through the crystal.  The latter component produces what are known as Neganov-Luke phonons~\cite{Luke}, which are proportional to both the collection voltage applied across the detector and the number of electron-hole excitations produced (i.e. the ionization signal). In order to keep the ionization and phonon measurements independent, one is limited to apply very low collection voltages across the detector ($\sim$Volts)  compared to the standard semiconductor detectors operating at  77 K ($\sim$1000 Volts). Although charge can be fully collected at such low fields for the events well within the bulk of the detector, for the events very close to the ionization electrodes a fraction of the carriers can back diffuse toward a wrong electrode (or to be trapped on the surface states).This results in an incomplete collection of charge. The ionization signal can be suppressed to the extent that, for an event occurring very close to the surface of the detector ($e.g.$ for a beta interaction), an ER can appear in the NR region and mimic a WIMP interaction. The solution incorporated by CDMS~II was to utilize the timing information of athermal phonons to reconstruct the location of events in the detector. Thus, CDMS II was able to identify and reject with $> 99.8\%$ efficiency those near surface events. Both CDMS and EDELWEISS are currently using a new technique based on a new interleaved ionization electrode geometry and a nonuniform tangential field at the surfaces of the detector to discriminate the near surface events by the ionization collection asymmetry between electrodes with excellent rejection efficiencies~\cite{iZIP}$^{,}$~\cite{ID}.

Running with 10$\times$400 g Interdigitated germanium Detectors (ID) and accumulating an exposure of 427 kg.d at the Modane underground laboratory (1800 m deep), EDELWEISS II observed 5 candidate events in the range of [20-200 keV] consistent with an expected background of 3~\cite{Edelweiss2011}. EDELWEISS III will deploy 40 kg of larger Fully Interdigitated Detectors (FID 800 g) with improved ionization collection geometry to also cover the cylindrical surfaces of the detectors with the expected sensitivity of ~$\sigma_{SI}\sim1\times10^{-45} cm^{2}$. 

The CDMS~II setup at the Soudan underground laboratory in Minnesota, US (713 m deep),  consisted of 4.5 kg (1.3 kg) of Ge (Si) Z-sensitive Ionization and athermal Phonon (ZIPs)  detectors 250 g(106 g) each. The final results from CDMS~II Ge exposure for  energy interval [10-100 keV] and low threshold (2 keV) is shown in Fig.~\ref{fig:SILimPlot} and described in detail elsewhere~\cite{c58science}$^{,}$~\cite{SoudanLowthresh2011}. Both CDMS II and EDELWEISS use Ge as their target absorber and performed a combined analysis of the two collaborations' WIMP exclusion region which resulted in slightly lower limit at the high mass but incurred a penalty at lower mass due to un-rejected backgrounds~\cite{CDMS_Edelweiss}. CDMS~II recently published results of the 142 kg.d exposure from operation of 11 Si ZIPs~\cite{CDMSIISi}. After all the analysis cuts, 3 events have been observed inconsistent with the expected background of 0.7 event. A likelihood analysis with the CDMS known backgrounds was performed and the highest likelihood occurs for WIMP mass of 8.6 GeV/$c^{2}$ and a cross section of $1.9\times10^{-41} cm^{2}$. The results are consistent with CDMS previous low threshold and EDELWEISS II low threshold analysis but in strong tension with XENON10 and XENON100 results~\cite{Xenon100final}.  Recent studies show that those inconsistencies can be softened with different assumption for the relative scintillation yield in XENON~100~\cite{Hooper Xenon}. The LUX experiment~\cite{LUX}  (currently running) with a significantly higher light yield than XENON~100 and SuperCDMS Soudan can confirm or further constrain the WIMP hypothesis in this mass region.

SuperCDMS Soudan is currently running with 9 kg of recently designed $\sim$600 g interleaved ZIP (iZIP) Ge detectors~\cite{iZIP} to reach WIMP-nucleon sensitivity of $\sigma_{SI}\sim3\times10^{-45} cm^{2}$ for WIMPs mass of $\sim$60 GeV/c$^{2}$. The background rejection performance of the CDMS new iZIP design was studied and shown to be far better than what is required for SuperCDMS Soudan sensitivity goals. The same design is adapted as the baseline for future generation of the experiment: SuperCDMS SNOLAB~\cite{SuperCDMS SNOLAB}. SuperCDMS SNOLAB is currently under R$\&$D to deploy 200 kg of 1.5 kg iZIP detector modules in SNOLAB underground laboratory in Ontario, Canada (2070 m deep). The goal of the experiment is to achieve 1 background leakage over 4 years of operations and to reach WIMP-nucleon sensitivity of $\sigma_{SI}\sim1\times10^{-46} cm^{2}$.

\begin{figure*}[htbp]
\vspace{-10pt}
\begin{center}
\centering
\includegraphics[  width=0.6\linewidth,  keepaspectratio]{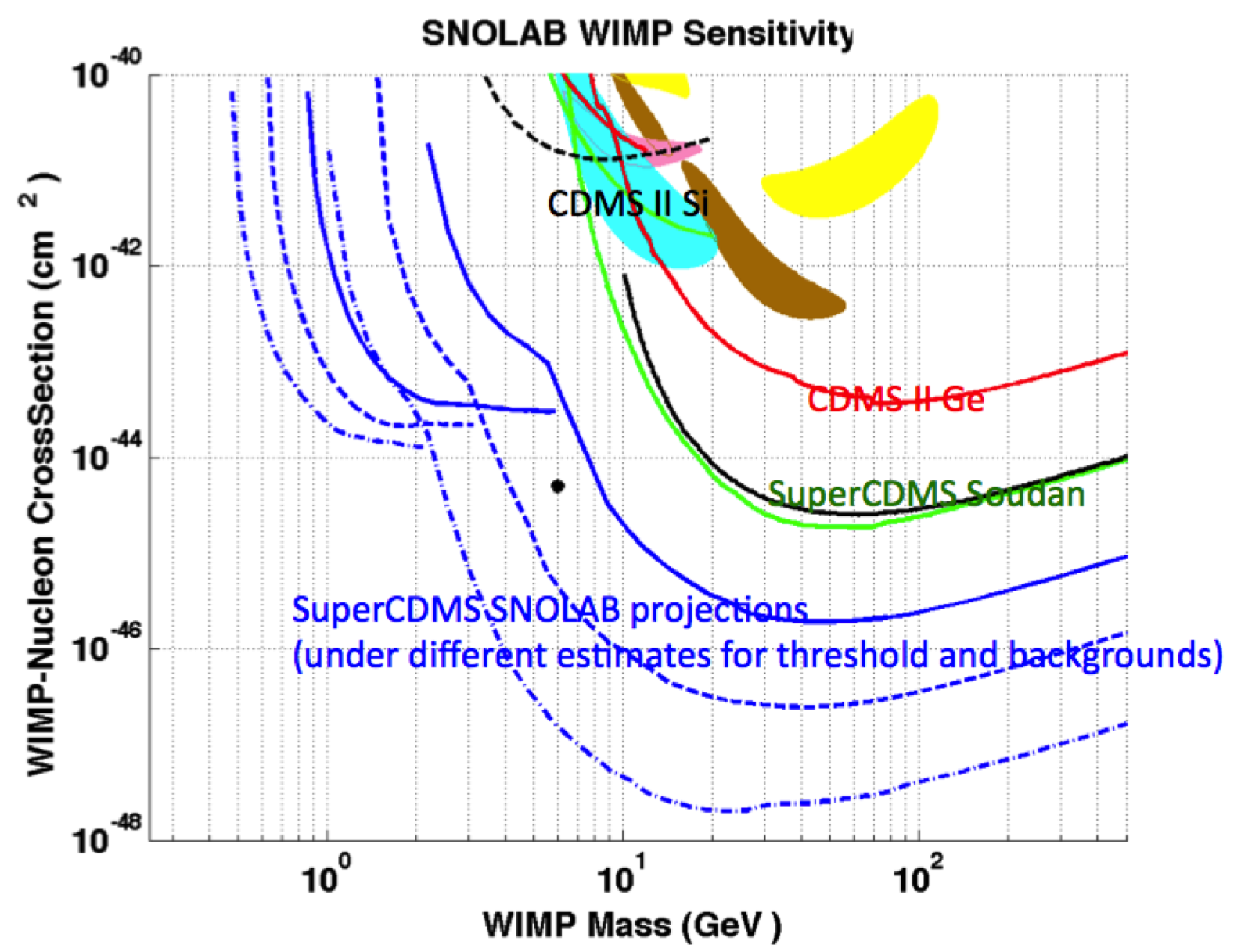}
\caption{\footnotesize WIMP-nucleon cross section exclusion limits as a function of WIMP mass for Xenon 100 (green) and CDMS~II Ge (red) as well as the 90 $\%$ CL signal contours for the low energy event excesses found in CDMS~II Si (cyan), CRESST(brown), CoGENT (pink), and DAMA (yellow).  5 (solid blue) and 10 year (dashed blue) sensitivity projections for the proposed 200kg SuperCDMS experiment at SNOLAB and a 1 ton upgrade with improved performance (dash-dot blue left curves) are also shown.  In each case, CDMS detectors operating with a 100V bias for significant Luke-Neganov gain have the greatest reach at low masses. The $^{8}$B solar neutrino coherent interaction in SuperCDMS Ge detectors can produce very similar spectrum as WIMPs of mass $\sim$7 GeV/c$^{2}$ and cross-section$\sim$10$^{-45}$ cm$^{2}$ is also shown on this plot (black dot)}
\label{fig:SNOLAB Low threshold}
\end{center}
\vspace{-10pt}
\end{figure*}

Next generation light mass dark matter searches are strongly motivated by both the near threshold signal excesses found in DAMA~\cite{HooperPRD2010} , CRESST~\cite{CRESSTII730kgd}, and CoGENT~\cite{CoGeNT} and a large number of natural theoretical models. Unfortunately, the energy resolution requirements of potential next generation dark matter experiments are over an order of magnitude lower than found in the current generation of massive detector cryogenic experiments. Thus, a focused research and development effort focused on substantially improving energy thresholds is required. CDMSlite is a CDMS chid project that uses the Neganov-Luke phonon amplification by operating detectors at very large ionization bias ($\sim$~70 Volts) to effectively measure the ionization via phonons with an excellent S/N $\sim$15 eV$_{ee}$. The analyses of the data form CDMSlite with very low energy threshold is being finalized and will soon be available to public. In the absence of large current leakage the Signal-to-Noise (S/N) and thus the detection threshold can further be improved to $\sim$1 eV by applying even larger biases. This goal should be achievable since the standard Ge gamma detectors at 77 K are regularly operated at biases $\sim$1000 Volts. 

Another approach would be to optimize the phonon sensors for better resolution by operating the detectors at lower 
base temperatures using superconducting Transition Edge Sensors (TES) with lower T$_{c}$'s. It can be shown~
\cite{Matt thesis} that the energy sensitivity in CDMS detectors scales as $T_{c}^{-3}$. CDMS intends to 
experimentally validate this scaling law and to achieve the $\sim$eV resolution needed for an ultimate light mass 
dark matter detection. Fig.~\ref{fig:SNOLAB Low threshold} shows the possible sensitivity reach for SuperCDMS 
SNOLAB experiment at the low mass region of WIMP parameter space using a combination of methods described 
above.
 
\section{Conclusion}

The direct detection of dark matter is a challenging goal, requiring detectors with both excellent background rejection and a very low energy threshold. Low temperature detectors are unique in that they offer a low threshold as well as event-by-event background rejection. This enables the exploration of a wide range of hypothetical WIMP masses down to $<$GeV/c$^{2}$. The ongoing experiments described here all plan to increase their background rejection efficiencies commensurate to their increased WIMP exposure. This insures a linear increase in WIMP detection sensitivity with the accrued exposure ($MT$) rather than $\sqrt{MT}$ of the experiments limited by background. Regarding future experiments, SuperCDMS, SNOLAB, and EURECA (a joint effort by both EDELWEISS and CRESST) are aiming to reach WIMP-nucleon sensitivity of $\sim$10$^{-46}$ cm$^{2}$.

\end{document}